\documentclass[a4,10pt,twoside]{article}

\usepackage[utf8]{inputenc}
\usepackage[english]{babel}
\usepackage[T1]{fontenc}
\usepackage{url}
\usepackage[colorlinks=true, allcolors=blue]{hyperref}
\usepackage{graphicx}
\usepackage[left=85pt, right=85pt, top=85pt, bottom=95pt,
headheight=15pt,%
headsep=10pt]{geometry}%
\usepackage{amsmath, amssymb}
\usepackage{authblk}
\usepackage{doi}
\usepackage{newtxtext,newtxmath}
\usepackage{microtype}
\usepackage{booktabs}
\usepackage[bf,small]{caption}

\usepackage[figurename=Fig.]{caption}

\title{Maximally localized modes of a multimode fiber}

\author[1,*]{Nicolas Barré}

\affil[1]{Independent researcher}

\affil[*]{email: nicolas.barre@protonmail.com}

\date{\today}

\begin{document}

\maketitle

\begin{abstract}
  This article presents an optimization method to find the most spatially concentrated basis
  of a multimode fiber, obtained by minimizing the sum of the spatial spreads of the
  individual modes over all unitary transformations of a given orthonormal mode set. The
  resulting modes are the optical analogue of maximally localized Wannier functions in
  solid-state physics. We apply the method to the Laguerre-Gaussian basis of a graded-index
  fiber for mode counts ranging from 6 to 55. In all cases, the modes spontaneously organize
  into concentric rings without any geometric constraint being imposed. The spot sizes and
  ellipticities evolve from one ring to the next in ways that geometric packing approaches
  cannot predict. For large mode counts, the optimizer finds solutions where neither the
  number of spots per ring nor the spots within a given ring follow a regular pattern,
  indicating that the fully symmetric arrangement is no longer a minimum of the spread
  functional. A constrained variant of the method enables the optimizer to target any
  prescribed bundle geometry while quantifying its localization cost, opening a route to
  physically grounded photonic lantern design.
\end{abstract}

\section{Introduction}

Photonic lanterns are adiabatically tapered waveguide structures that provide a low-loss
interface between a multimode fiber and a bundle of single-mode
fibers~\cite{Birks:2015}. They have become enabling components in two distinct domains:
space-division multiplexing (SDM)~\cite{Velázquez-Benítez:2018}, where they serve as mode
multiplexers coupling individual single-mode channels to the spatial modes of a few-mode
fiber, and astrophotonics, where they reformat multimode telescope light into single-mode
outputs compatible with high-resolution photonic instrumentation~\cite{Jovanovic:2023}.

In both contexts, the spatial arrangement of the single-mode fiber cores is a critical
design parameter. Fontaine et al.~\cite{Fontaine:2012} established systematic geometric
requirements for efficient mode conversion, showing that one ring of fibers per radial mode
group is needed, with the number of fibers per ring set by the azimuthal content. A
complementary approach treats the problem as a geometric optimization: Davenport et
al.~\cite{Davenport:2021}, building on dense circle packing results~\cite{Graham:1998},
optimized core arrangements by maximizing packing density, treating all fibers as identical
disks. Both approaches share a common assumption, namely that the bundle geometry is a
design input, either constrained by modal compatibility or by geometric optimality. Neither
asks what geometry the fiber modes themselves prefer.

In this work we address this question directly. Given a basis of $N$ orthonormal modes of a
multimode fiber, we seek the unitary transformation that yields the most spatially
concentrated basis, minimizing the total spatial spread across all modes. This problem is
the optical analogue of maximally localized Wannier functions~\cite{Wannier:37, Marzari:97}:
in that framework, Bloch eigenstates are unitarily transformed into localized orbital-like
functions by minimizing a spread functional. Our problem is the non-periodic, single
$k$-point limit of this construction, where a single unitary matrix replaces the
$k$-dependent gauge freedom.

The resulting modes, which we call Maximally Localized Fiber Modes (MLFM), reveal the
intrinsic spatial structure of the fiber without any geometric assumption. Applied to a
rotationally symmetric graded-index fiber, the method uncovers a rich phenomenology: the
modes self-organize into concentric rings, the spot geometry evolves systematically across
rings in ways that circle packing cannot anticipate, and beyond a critical mode count, the
ring counts deviate from any regular progression, the spots within a given ring begin to
exhibit subtle geometric differences, and multiple topologically distinct solutions with
near-identical spread values coexist as stable local minima. We further show that a
constrained variant of the optimization, which penalizes deviations from a target bundle
geometry, recovers near-optimal symmetric solutions whose total spread remains within a few
percent of the unconstrained optimum, establishing a quantitative bridge between modal
physics and lantern design.

\section{Maximally localized fiber modes}

The modes of a multimode fiber form an orthonormal basis for the transverse fields that can
propagate within it, but this basis is not unique. Any unitary transformation of an
orthonormal set yields another equally valid orthonormal set. Among all such bases, we seek
the one whose modes are most spatially concentrated, minimizing the total spatial spread.
As we show below, this problem has a natural solution in terms of a matrix optimization over
the unitary group $\mathrm{U}(N)$, and reveals the intrinsic spatial structure of the fiber
modes without any geometric assumption. We refer to the resulting modes as Maximally
Localized Fiber Modes (MLFM).

Let $\{\varphi_i\}_{i=1}^N$ be an orthonormal basis of $N$ modes of a multimode fiber,
defined on a two-dimensional transverse grid. We seek a unitary transformation
$\mathbf{U} \in \mathrm{U}(N)$ yielding a new basis
\begin{equation}
    \psi_i = \sum_j U_{ij}\, \varphi_j,
\end{equation}
whose total spatial spread is minimal. Following~\cite{Marzari:97}, we define the
spread functional
\begin{equation}
    \Omega(\mathbf{U}) = \sum_{i=1}^N \sigma_i^2,
    \qquad
    \sigma_i^2 = 2\left(\langle x^2 \rangle_i - \langle x \rangle_i^2
                       +\langle y^2 \rangle_i - \langle y \rangle_i^2\right),
\end{equation}
where $\langle \cdot \rangle_i$ denotes the intensity-weighted spatial average over mode
$\psi_i$. The factor of 2 is chosen such that $\sigma_i = w_0$ for the fundamental Gaussian
mode of waist $w_0$. This is the optical analogue of the Marzari--Vanderbilt functional for
maximally localized Wannier functions~\cite{Marzari:97, Marzari:2012}, reduced to the single
$k$-point case appropriate for a non-periodic system.

Note that $\Omega$ is invariant under any global rotation of the transverse plane, so the
minimizer is defined only up to an overall rotation of the mode pattern. When the input
basis has rotational symmetry, which is the case for any basis whose modes can be labeled by
a radial index $p$ and an azimuthal index $l$, this invariance has a precise algebraic
form. Let $\mathbf{U}_{\theta_0}$ be the diagonal matrix with entries $e^{il_i\theta_0}$,
where $l_i$ is the azimuthal index of mode $\varphi_i$.  Since each mode carries an
azimuthal phase $e^{il_i\theta}$, multiplying by $e^{il_i\theta_0}$ shifts the angular
coordinate by $\theta_0$ consistently across all modes, which amounts to rotating the input
basis by $\theta_0$.  As a consequence, if $\mathbf{U}$ is a minimizer of $\Omega$, then
$\mathbf{U} \cdot \mathbf{U}_{\theta_0}$ is also a minimizer for any $\theta_0$, producing a
rotated version of the localized basis. The angle $\theta_0$ is therefore a free parameter
of the rotationally symmetric problem, and the optimization will select a particular value
depending on the initial condition.

To optimize over $\mathrm{U}(N)$ without manifold constraints, we parameterize
\begin{equation}
    \mathbf{U} = \exp\!\left(\frac{\mathbf{A} - \mathbf{A}^\dagger}{2}\right),
\end{equation}
where $\mathbf{A} \in \mathbb{C}^{N\times N}$ is unconstrained. The skew-Hermitian matrix
$(\mathbf{A} - \mathbf{A}^\dagger)/2$ is an element of the Lie algebra $\mathfrak{u}(N)$,
guaranteeing that $\mathbf{U}$ remains unitary throughout optimization. The optimization is
implemented in FluxOptics.jl~\cite{Barré:2026}, a differentiable wave optics framework that
provides mode generation, unitary transformations, and automatic differentiation. Parameters
are updated using Nesterov-accelerated gradient descent~\cite{Nesterov:1983}.

The unconstrained minimization of $\Omega$ reveals the intrinsic optimal
geometry of the fiber. When a specific target geometry is desired instead,
one can steer the solution by augmenting the functional with a quadratic
penalty on the mode centroids:
\begin{equation}
  \label{eq:omega_penalty}
    \Omega_\beta(\mathbf{U}) = \Omega(\mathbf{U})
    + \beta \sum_{i=1}^N
      \left[\left(\langle x \rangle_i - x_i^0\right)^2
           +\left(\langle y \rangle_i - y_i^0\right)^2\right],
\end{equation}
where $\{(x_i^0, y_i^0)\}$ is the set of target centroid positions defining the desired
bundle geometry, and $\beta \geq 0$ controls the trade-off between localization and geometry
fidelity. Setting $\beta = 0$ recovers unconstrained minimization, whereas increasing
$\beta$ constrains the centroids toward the target positions while still minimizing the
residual spread. This provides a principled tool for inverse design: given any candidate
bundle geometry, one can quantify the localization cost of imposing it.

\section{Results and Discussion}

We illustrate the unitary localization algorithm on the Laguerre-Gaussian (LG) basis of a
graded-index fiber, which serves as a natural and representative case study. The method
itself is basis-independent and applies to any modal family. We restrict the study to
complete mode groups, giving mode counts $N \in \{6, 10, 15, 21, 28, 36, 45, 55\}$
corresponding to groups up to order 2 (for $N = 6$) through order 9 (for $N = 55$). This
restriction is motivated both physically and combinatorially. Physically, incomplete mode
groups tend to produce modes without a well-defined single dominant lobe, making the bundle
geometry interpretation less meaningful. Combinatorially, complete mode groups correspond to
triangular numbers $T_m = m(m+1)/2$, and this structure imposes a strong constraint on the
possible ring geometries, as we now show.

For a rotationally symmetric input basis, one expects the localized modes to arrange
themselves in concentric rings, a structure we now characterize combinatorially. More
precisely, denote by $n_k$ the number of spots in ring $k$, with $k = 1$ the innermost ring,
and let $S_K = \sum_{k=1}^K n_k$ be the cumulative spot count after $K$ rings. For the
bundle to correspond to a complete set of mode groups at each level, $S_K$ must be
triangular for every $K$. Under the linear recurrence $n_k = n_{k-1} + 4$, one can show that
the only two initial conditions satisfying this constraint for all $K$ are $n_1 = 1$, giving
$S_K = T_{2K-1}$, and $n_1 = 3$, giving $S_K = T_{2K}$. Together they cover all triangular
numbers, and any other starting value fails at some $K$. These two families are therefore
the only regular concentric ring geometries compatible with complete mode groups at every
level of the hierarchy. In the following, this recurrence relation will serve as a reference
geometry, and will be used to define regular target geometries to introduce as penalties in
$\Omega_\beta$~\eqref{eq:omega_penalty} when a symmetric solution is not naturally found.

\begin{figure}[!ht]
  \centering
  \includegraphics[scale=1]{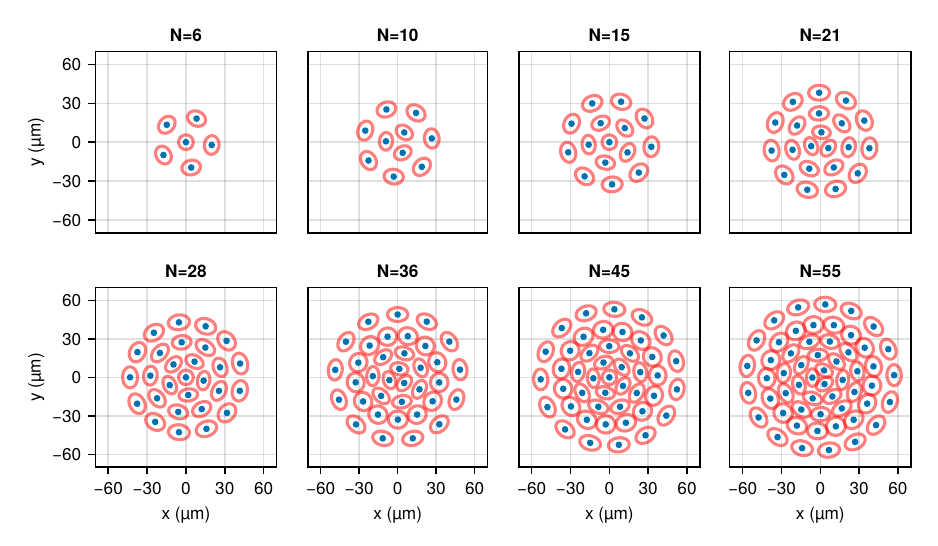}
  \caption{MLFM obtained by minimizing $\Omega$ over the LG basis of a graded-index fiber,
    for $N = 6$ to $55$. Each spot is represented by its intensity centroid (dot) and an
    ellipse whose semi-axes are proportional to $\sigma_r$ and $\sigma_\theta$ (scaled by
    0.45 for clarity). The two families of solutions ($n_1 = 1$ and $n_1 = 3$) are clearly
    visible.}
  \label{fig:bundles}
\end{figure}

For each $N$, we run 20 independent optimizations from random initializations of
$\mathbf{A}$. In most cases all runs converge to the same ring topology; exceptions are
$N = 28$, where $[1, 5, 11, 11]$ appears in 2 out of 20 runs alongside the dominant
$[1, 5, 9, 13]$, and $N = 55$, where $[3, 7, 11, 17, 17]$ and $[3, 7, 13, 15, 17]$ appear
with equal frequency. In all cases we retain and display the solution with lowest $\Omega$
in Fig.~\ref{fig:bundles}. The most striking result is that the modes spontaneously organize
into concentric rings. No geometric constraint of any kind is imposed, and the ring
structure emerges purely from the minimization of $\Omega$. The observed ring counts follow
the two families described above for $N \leq 28$, with $n_1 = 1$ for $N = 6, 15, 28$ and
$n_1 = 3$ for $N = 10, 21$, consistently following $n_k = n_{k-1} + 4$. For $N \geq 36$, the
initial condition $n_1 \in \{1, 3\}$ is preserved but the regular progression breaks down
beyond a certain ring: the optimizer finds $[3, 7, 13, 13]$, $[1, 5, 9, 15, 15]$, and
$[3, 7, 11, 17, 17]$ for $N = 36$, $45$, and $55$ respectively. These asymmetric
configurations will be revisited in the context of constrained optimization below.

Unlike the identical disks assumed in geometric packing approaches~\cite{Davenport:2021,
  Graham:1998}, the MLFM spots are heterogeneous. Both $\sigma^2_r$ and $\sigma^2_\theta$
grow with ring index: outer spots are larger and more elliptical than inner ones, and spots
are systematically more extended in the azimuthal direction than in the radial direction
($\sigma^2_\theta > \sigma^2_r$), reflecting the curvature of the ring geometry. The ring
spacing is also non-uniform and depends on the choice of input basis, reflecting the radial
structure of the modes. These are direct predictions of the localization functional,
invisible to any purely geometric argument.

\begin{figure}[!ht]
  \centering
  \includegraphics[scale=1]{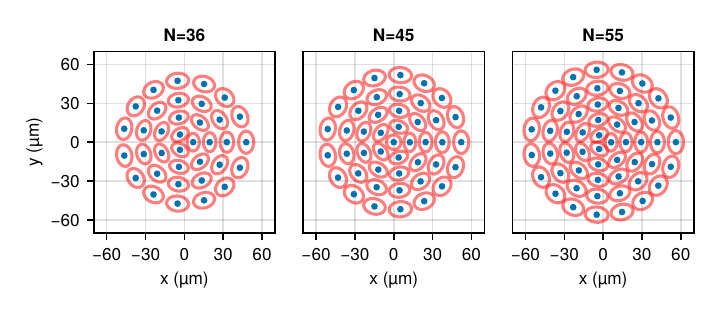}
  \caption{MLFM obtained with constrained optimization ($\beta = 1$) targeting
  the regular concentric ring geometry, for $N = 36$, $45$, and $55$.
  The total spread is within $0.8\%$, $1.4\%$, and $1.7\%$ of the
  unconstrained optimum respectively, confirming that the regular packing is
  near-optimal.}
  \label{fig:constrained}
\end{figure}

For $N \geq 36$, the unconstrained optimizer finds asymmetric solutions. To investigate
whether symmetric solutions exist and at what cost in total spread, we apply the constrained
optimization with $\beta = 1$, using target bundle geometries generated from the recurrence
rule, with $n_1 = 3$ for $N = 36$ and $N = 55$, and $n_1 = 1$ for $N = 45$, with spots
distributed uniformly on each ring. The results are shown in Fig.~\ref{fig:constrained}.

The constrained optimization successfully recovers symmetric solutions in all three cases.
Interestingly, the centroids of the resulting modes do not coincide with the target
positions specified in the constraint. The optimizer preserves the angular symmetry of the
target geometry but freely adjusts the ring radii, showing that the method is robust to an
imprecise specification of the target geometry. One does not need prior knowledge of the
optimal ring radii for the constraint to produce a well-organized symmetric solution.
 
The total spread of the constrained solutions deviates by only $0.8\%$, $1.4\%$, and $1.7\%$
from the unconstrained optimum for $N = 36$, $45$, and $55$ respectively, confirming that
the regular packing geometry is near-optimal even when it is not the global minimizer of
$\Omega$. However, these symmetric solutions are not local minima of the unconstrained
problem. Releasing the constraint and resuming the optimization always leads away from the
symmetric configuration and recovers solutions of the type shown in Fig.~\ref{fig:bundles}.

A subtlety of the constrained solutions is that they are only \emph{quasi}-symmetric. The
target geometries used above were constructed by aligning the first spot of each ring along
the horizontal axis. One can instead introduce an arbitrary angular offset per ring in the
target positions and verify that the optimizer still converges to a solution with the same
angular structure, adjusting only the ring radii. However, the total spread of the resulting
solution varies slightly (by $\sim 0.05\%$) depending on the chosen offsets. This
sensitivity would not occur if the problem had a true continuous rotational symmetry, and
reveals that the constrained solutions are only quasi-symmetric: spots within the same ring
are not exactly identical, even though their intensity profiles are visually hardly
distinguishable.
 
\begin{figure}[!ht]
  \centering
  \includegraphics[scale=1]{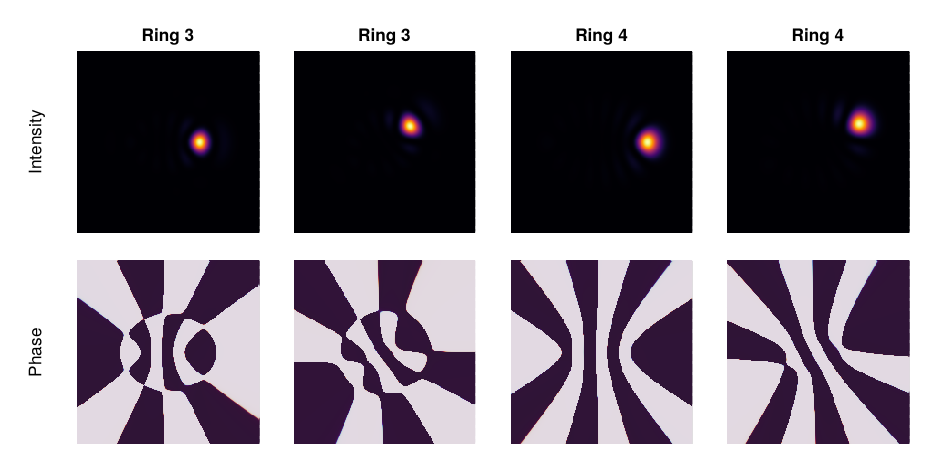}
  \caption{Intensity (top) and phase (bottom) of representative MLFM for the
  $N = 36$ constrained solution. Columns 1--2: two adjacent spots from ring 3;
  columns 3--4: two adjacent spots from ring 4. Within each ring, the dominant
  intensity lobes are nearly identical while the nodal line patterns differ,
  illustrating the quasi-symmetry of the constrained solutions. Modes from
  different rings have partially overlapping spatial supports; orthogonality
  is ensured by the phase structure.}
  \label{fig:modes}
\end{figure}

This quasi-symmetry is illustrated in Fig.~\ref{fig:modes}, which shows the intensity and
phase of two representative modes from ring 3 and two from ring 4 for the $N = 36$
constrained solution. Within each ring, the dominant intensity lobes are nearly
indistinguishable, yet the nodal line patterns in the phase differ from one spot to the
next. It is precisely this phase structure that ensures orthogonality between modes whose
intensity lobes partially overlap: although neighboring modes share significant spatial
support, the overlap integrals vanish exactly as required by the unitarity of the
transformation.

Unlike geometric packing approaches~\cite{Graham:1998, Davenport:2021}, which
optimize the arrangement of identical disks independently of the modal content
of the fiber, the present method derives the bundle geometry directly from the
modes. Beyond geometry identification, the constrained functional $\Omega_\beta$
provides a quantitative measure of how well any prescribed bundle geometry is
matched by the fiber modes, which may serve as a criterion for comparing
candidate lantern designs.

\section{Conclusion}

We have introduced an optimization method to find the most spatially concentrated basis of a
multimode fiber by minimizing a spread functional over all unitary transformations of a
given orthonormal mode set. Applied to the Laguerre-Gaussian basis of a graded-index fiber,
the method reveals that the optimal modes spontaneously organize into concentric rings whose
structure is entirely determined by the modal content of the fiber, without any geometric
assumption. The spot geometry evolves systematically across rings, and beyond a critical
mode count the regular recurrence $n_k = n_{k-1} + 4$ no longer holds: the ring counts
deviate from this rule and spots within the same ring are no longer identical in shape,
exhibiting different radial and azimuthal variances. These are genuine properties of the LG
basis at large mode counts, not artifacts of the optimization. A constrained variant of the
method recovers near-optimal symmetric solutions at the expense of a small additional spread,
and shows that symmetric configurations are not local minima of the unconstrained problem.

These results establish a direct link between the modal structure of a multimode fiber and
the optimal geometry of a photonic lantern bundle, complementing existing approaches based
on modal compatibility constraints~\cite{Fontaine:2012} or geometric circle
packing~\cite{Davenport:2021}. The constrained functional $\Omega_\beta$ further provides a
quantitative tool for comparing candidate bundle geometries in terms of their localization
cost, providing a physically motivated criterion for evaluating and optimizing photonic
lantern bundle designs. Extensions to other modal bases or fiber geometries are
straightforward within the same framework.

\paragraph{Funding.}
This research received no external funding.

\paragraph{Data availability.}
All simulations were performed using FluxOptics.jl~\cite{Barré:2026}, an open-source Julia
package for differentiable wave optics. Data and code underlying the results presented in
this paper are available from the author upon reasonable request.

\let\OLDthebibliography\thebibliography
\renewcommand\thebibliography[1]{
  \OLDthebibliography{#1}
  \setlength{\parskip}{1ex}
  \setlength{\itemsep}{0pt plus 1ex}
}

\bibliographystyle{unsrturl}
\bibliography{localized_modes}

\end{document}